\documentclass[11pt]{article}
\usepackage{epsfig}
 \csname
@addtoreset\endcsname{equation}{section}

\newcommand{\hs}[1]{\hspace{#1 mm}}

\newcommand{\ba}{\begin{eqnarray}}
\newcommand{\ea}{\end{eqnarray}}
 
\newcommand{\ben}{\begin{displaymath}}
\newcommand{\een}{\end{displaymath}}

\newcommand{\ft}[2]{{\textstyle {\frac{#1}{#2}} }}

\def\sqr#1#2{{\vcenter{\vbox{\hrule height.#2pt
         \hbox{\vrule width.#2pt height#1pt \kern#1pt
            \vrule width.#2pt}
         \hrule height.#2pt}}}}
\def\square{\mathop{\mathchoice\sqr56\sqr56\sqr{3.75}4\sqr34\,}\nolimits}

\def\a{\alpha}
\def\b{\beta}
\def\c{\gamma}

\def\d{\delta}
\def\D{\Delta}
\def\e{\epsilon}

\def\f{\phi}

\def\k{\kappa}
\def\l{\lambda}

\def\m{\mu}

\def\o{\omega}

\def\del{\partial}
\newcommand{\la}[1]{\label{#1}}
\let\bm=\bibitem
\def\nn{\nonumber}
\newcommand{\eq}[1]{(\ref{#1})}

\newcommand{\w}[1]{\\[0.#1cm]}

\def\be{\begin{equation}}
\def\ee{\end{equation}}
\def\bea{\begin{eqnarray}}
\def\eea{\end{eqnarray}}

\def\ft#1#2{{\textstyle{{\scriptstyle #1}
\over {\scriptstyle #2}}}}

\def\vare{\varepsilon}

\def\ed{\end{document}}

\newcommand{\tamphys}{Feza Gursey Institute, Cengelkoy 81220,
Istanbul, Turkey}

\newcommand{\auth}{\large 
Nihat Sadik Deger \footnote{deger@gursey.gov.tr}}

\begin{document}

\vspace{20pt}

\hfill{\today}

\begin{center}

{\Large \bf Renormalization Group Flows from D=3, N=2}

{\Large \bf Matter Coupled Gauged Supergravities}

\vspace{30pt}


\auth

\vspace{15pt}

\tamphys

\vspace{30pt}

{\bf Abstract}

\end{center}

We study holographic RG flows of $N=2$ matter coupled
$AdS_3$ supergravities which admit both compact and 
non-compact sigma manifolds. For the
compact case the supersymmetric domain wall solution interpolates between 
a conformal 
IR region and flat spacetime and this corresponds to a deformation of 
the CFT by an irrelevant operator.
When it
is non-compact, the solution can be interpreted
as a flow between an UV fixed point and a non-conformal(singular) IR 
region. This is an exact example of a deformation 
flow when the singularity is physical.
We also find a non-supersymmetric deformation flow when 
the scalar 
potential has a second $AdS$ vacua. The ratio of the central charges is 
rational for certain values of the size of the sigma model. 
Next, we analyze the 
spectrum of a massless scalar on our background by transforming the 
problem into Schr\"{o}dinger form. 
The spectrum is continuous for the 
compact case, yet it can be both continuous (with or
without mass gap) and discrete otherwise. 
Finally, 2-point functions are computed for two examples whose 
quantum mechanical potentials are of Calogero type.

{\vfill\leftline{}\vfill

\pagebreak

\setcounter{page}{1}


\section{Introduction}


After the celebrated $AdS/CFT$ conjecture \cite{mal1,mal2,mal3}
renormalization group (RG) flows from gauged supergravities have been 
studied extensively (see \cite{rev1,rev2,rev3} for review and 
references). However, $D=3$ seems to 
be an exception although it 
has the obvious advantage of being dual to a 2-dimensional CFT. 
To the best of our knowledge only papers where RG flows are worked in 
detail using 3D-supergravities are \cite{sam2,sam3} in which
an exact RG flow from $N=8$
model with $SO(4) \times SO$(4) gauge symmetry was found and its 
correlation functions were analyzed. 

One reason for this neglect might be that,  
matter coupled $AdS_3$
supergravities are not as familiar as their higher dimensional relatives 
and their construction is still in progress.  
Therefore, we would like to begin with a brief summary of the current 
status in
$D=3$. The $N=2$ supergravity 
coupled to an arbitrary number of scalar
supermultiplets was constructed in \cite{it, ads2}. 
In \cite{ads2} scalars are charged under the  U(1) R-symmetry group 
and consequently they have a potential, whereas in \cite{it} there 
is only a cosmological constant. The connection between these two models 
for the flat sigma model was described in \cite{ads3}. In \cite{ads2} the 
$N=1$ truncation was obtained too. Another $N=1$ model with a 
different 
sigma-model manifold was given in \cite{nishino1}. The boundary 
symmetries of \cite{ads2} were studied in 
\cite{ads3} and its extension
by including a Fayet-Iliopoulos term was given in
\cite{sam1}. The maximal ($N=16$) gauged supergravity
was constructed in
\cite{nic1, nic2} and its vacua were studied in 
\cite{fisch1, fisch2}. Recently a topological generalization of $N=16$ 
model was obtained in \cite{nishino2}.
The $N=8$  matter coupled $AdS_3$ supergravities 
were 
constructed in \cite{nic3}. 
There also has been some effort to obtain $AdS_3$ gauged supergravities 
directly 
by 
dimensional reduction \cite{pope1,pope2,pope3}.

The $N=2$ model that we will study here \cite{ads2} has the virtue of 
being simple and wealthy at the same time. It 
admits both compact and non-compact sigma model manifolds and it has three 
free 
parameters: Gravitational 
coupling constant, cosmological constant and the size of the 
sigma manifold which we denote by ``$a$".
This extra freedom gives a richness since    
the value of ``$a$" changes the shape of the potential drastically when 
the scalar manifold is non-compact.

The organization of this paper is as follows; 
in the next section we will review the $N=2$ gauged supergravity and 
rederive the supersymmetric domain wall solution obtained in \cite{ads2} 
in a form which is more useful for our aims. 
In general, there is 
an ambiguity whether a solution is a true deformation or a different 
vacua of the same theory \cite{bkl,kw}. Our first goal is to resolve 
this for 
our model.
When the sigma manifold is compact the superpotential is a 
trigonometric function of the scalar field. Such a superpotential was 
studied in
\cite{behrndt} for $D=5, N=2$ gauged supergravity and our result
is similar too: The supersymmetric solution interpolates 
between a conformal IR region and flat spacetime.
This is created by deforming the dual CFT by an addition of an irrelevant 
operator.
In the non-compact 
case, the solution 
is asymptotically $AdS$ and it 
exhibits a naked
singularity. We will show that
this can be interpreted as a RG flow 
to a non-conformal IR theory \cite{gir}. When $a^2\leq 1/2$ this  
corresponds to a non-renormalizable scalar mode and therefore to a 
deformation of the CFT by an addition of a relevant operator. 
Whereas, 
for $a^2>1/2$ the flow is 
driven
by giving a nonzero 
expectation value 
to the operator. The singularity is unphysical for this case since the 
potential 
becomes unbounded \cite{gubser} in supersymmetric solution.
When $1/2 < a^2 < 1$ the potential has another (stable) $AdS$ extrema and 
one may 
try to construct a non-supersymmetric interpolating solution. 
However, an exact solution is 
hard to obtain. Instead we will follow the analysis done in 
\cite{seven} for 
$D=7, N=1$ gauged supergravity and find a numerical solution. 
This actually is  
enough to conclude that interpolating solution is a true deformation.
We also show that the ratio of the central charges and operator 
dimensions are rational for certain 
values of $a^2$.
In section 3, we solve the wave equation for a minimally coupled scalar in 
our 
background by casting the problem into Schr\"{o}dinger form.
The spectrum is continuous for the compact sigma manifold,   
while it can be continuous (there is a mass gap for $a^2=1/4$) or discrete 
otherwise, depending on $a^2$. 
Moreover, we compute the 2-point functions for
two specific values of $a^2$ when sigma manifold is non-compact. The 
quantum mechanical 
potentials for these, belong to the class of potentials that appear in 
Calogero 
models (3-body problem in 1-dimension) as was first observed in 
\cite{calogero} for $D=5, N=8$ gauged supergravity.
We conclude in 
section 4 with some comments and possible future directions.


\section{The Matter Coupled, $N=2 \ AdS_3$ Supergravity}


The $N=2\, AdS_3$ supergravity multiplet consists of a graviton
$e_\m{}^a$, a complex gravitini $ \psi_\m$ and a gauge field $A_\m$.
The
$N=2$ scalar multiplet, on the other hand,
contains $n$ complex scalar fields $\phi^\a$ and $n$
complex fermions $\l^r$.

In \cite{ads2}, the sigma model manifold $M$ was taken to be a coset
space of the form $G/H$ where $G$ can be compact or
non-compact and $H$ is the maximal compact subgroup of
$G$. In particular, the following cases are considered:

\be
M_+= {SU(n+1)\over SU(n)\times U(1)}\ , \quad\quad M_-=
{SU(n,1)\over SU(n)\times U(1)}\ .
\la{pm2}
\ee

Note that $U(1)$ is the $R$-symmetry group.
We define the parameter $\e=\pm$1 to indicate the manifolds 
$M_\pm$.
In this paper we consider the cases $S^2=SU(2)/U(1)$ and 
$H^2=SU(1,1)/U(1)$, i.e, $n=1$. 
For our purposes it is enough have a single real scalar, consequently we 
set $\phi=|\phi|$ and all other fields to zero. \footnote {This is 
consistent with the field equations given in \cite{ads2}. The vector field 
equation implies that when the vector field is set to zero then the 
scalar field has to be real.
The critical 
points of this truncated theory with real $\f$ are also critical points 
of the 
full potential given in \cite{ads2}. This has already been noted in 
\cite{ads2} and can 
be seen from the form of potential which  depends on the
magnitude of the scalars at least quadratically.} To simplify the 
Lagrangian given in \cite{ads2} we make a redefinition for the scalar
field as follows:
 
\be
2\f \equiv \begin{cases}
{{\rm tanh}(\frac{\hat \phi}{2}) \hs{5} \e=-1,\cr\cr
{\rm tan}(\frac{\hat \phi}{2}) \hs{7} \e=1}
\end{cases}
\ee

Then, we drop the hat for notational convenience, and the Lagrangian 
of \cite{ads2} becomes:

\be
e^{-1} {\cal L}= \ft14 R -{1\over 4a^2} \del_\m\phi \del^\m \phi -
V(\phi)\ ,
\la{lagrangian}
\ee

where the potential is given by (Figure 1):

\be
V= -2m^2 {\rm C^2}\left[(1+2\e a^2){\rm C^2} -2\e a^2 \right] .
\label{pot}
\ee

The function C is defined as:

\be
{\rm C}= \begin{cases}
{{\rm cosh}\,\phi \hs{5} \e=-1,\cr\cr
{\rm cos}\,\phi \hs{7} \e=1}
\end{cases}
\la{v1}
\ee

The constant $``a"$ is the characteristic
curvature of $M_\pm$ (e.g. $2a$ is the inverse radius in the case
of $M_+=S^2$). 
The gravitational coupling constant $\k$ has been
set equal to one and $-2m^2$ is the $AdS_3$ 
cosmological 
constant.
Unlike in a typical $AdS$ supergravity coupled to matter,
the constants $\k,a,m$ are not related to each other for
non-compact scalar manifolds, while $a^2$ is quantized in terms of
$\k$ in the compact case as ${\k ^2 \over a^2}=n$ , where $n$ is an 
integer \cite{ads2}.

The fermionic supersymmetry transformations of the model are: 

\bea
\d \psi_\m &=& \left(\del_\m +\ft14 \o_\m{}^{ab} \c_{ab}\right)\vare + {1 
\over 2}\, W\c_\m \vare\ ,
\nn\w2
\d\l &=& {1\over 2}\left(-\c^\m \del_\m \phi - a{\del W 
\over \del 
\phi}\, \right)\vare\, .
\label{fermionic}
\eea

Here $W$ is the superpotential and is given by:  
 
\be
W= -2\e m{\rm C^2}
\label{superpot}
\ee
 
This is defined up to an overall unimportant sign. 
The potential $V$ can be written in terms of the superpotential as:
 
\be
V = \frac{a^2}{4} {\left(\del W \over \del \phi \right)}^2
- {1\over 2 }W^2  \ ,
\label{potential}
\ee

Let us now consider the following domain wall ansatz for the metric:

\be
ds^2 = e^{2A(y)} (-dt^2+dx^2) + dy^2\ ,
\la{m}
\ee

The field equations of (\ref{lagrangian}) with the above metric are:
 
\bea
\label{linear}
\phi '' + 2A'\phi ' &=& 2a^2 {\del V \over \del \phi}  \\
A''+2(A')^2 &=&-4V \nn \\
2A'' + 2(A')^2&=&-{(\phi ')^2 \over a^2} -4V \ .\nn
\eea

One of the three equations is redundant and can be derived from the other 
two. 
These are nonlinear equations and it is hard to find an exact solution 
except when there is supersymmetry. 
After imposing the vanishing of the supersymmetry conditions $\d\l=0$ and 
$\d\psi=0$ one 
obtains the following first order equations:

\be
\phi'= a^2 {\del W \over \del\phi} \ ,
\label{susy}
\ee

\be
A' = - W\ .
\label{ap}
\ee

where the prime indicates differentiation with respect to $y$. In 
deriving these equations we also 
imposed the condition $\c^y\vare= \e \vare$
which means that the solution is half-supersymmetry preserving.
From $\d\psi=0$ the $\phi$-dependence of the spinor $\vare$ is 
determined to be \cite{ads2}:

\be
\vare = (\e - \e {\rm C^2})^{1/8a^2} \,(1 -\e \c_y)\vare_0 
\ee
where $\vare_0$ is an arbitrary constant spinor. 

\begin{figure}
\centerline{\epsfxsize=3.3truein
\epsffile{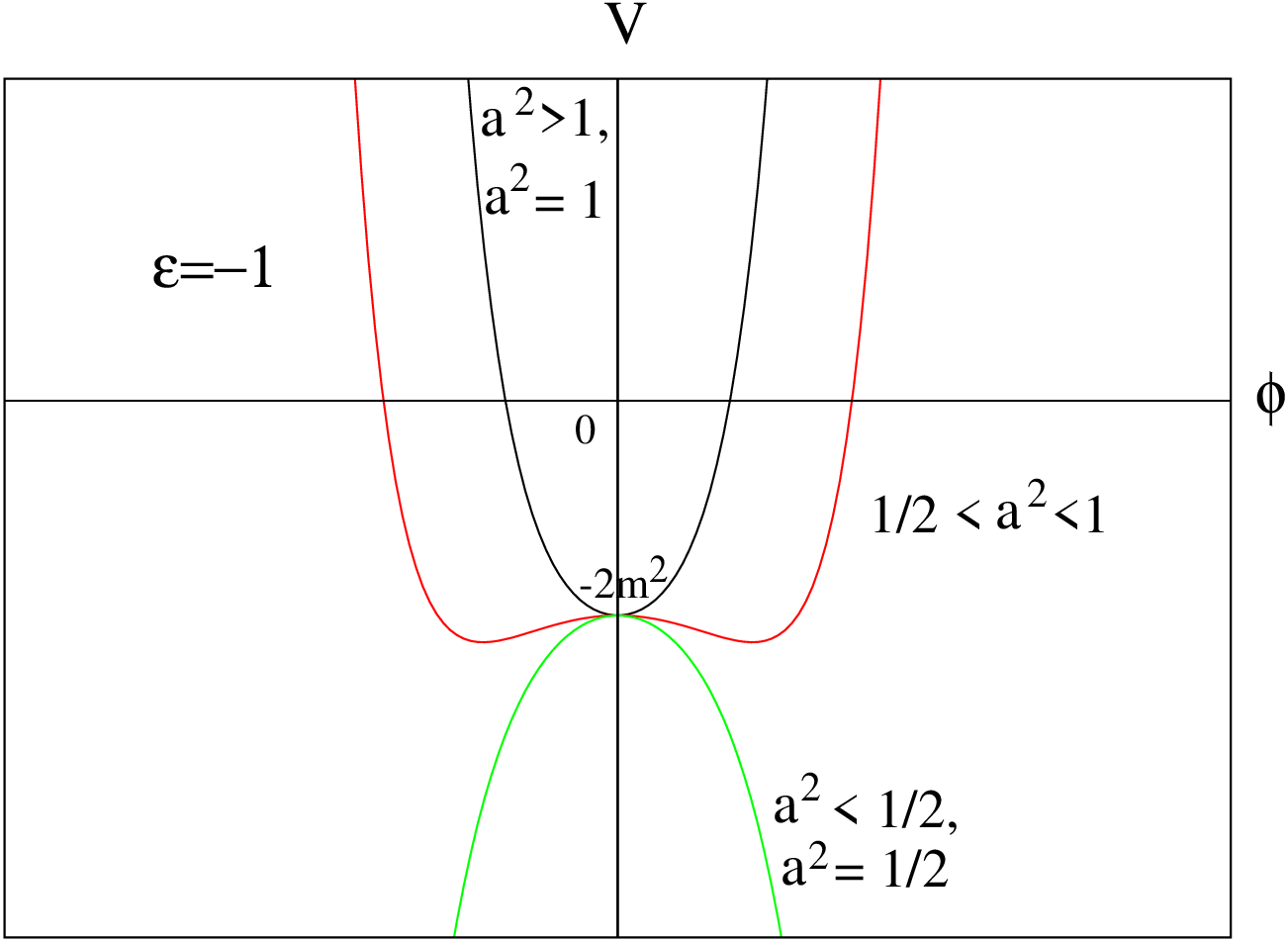}
\hspace{0.4in}
\epsfxsize=3.3truein
\epsffile{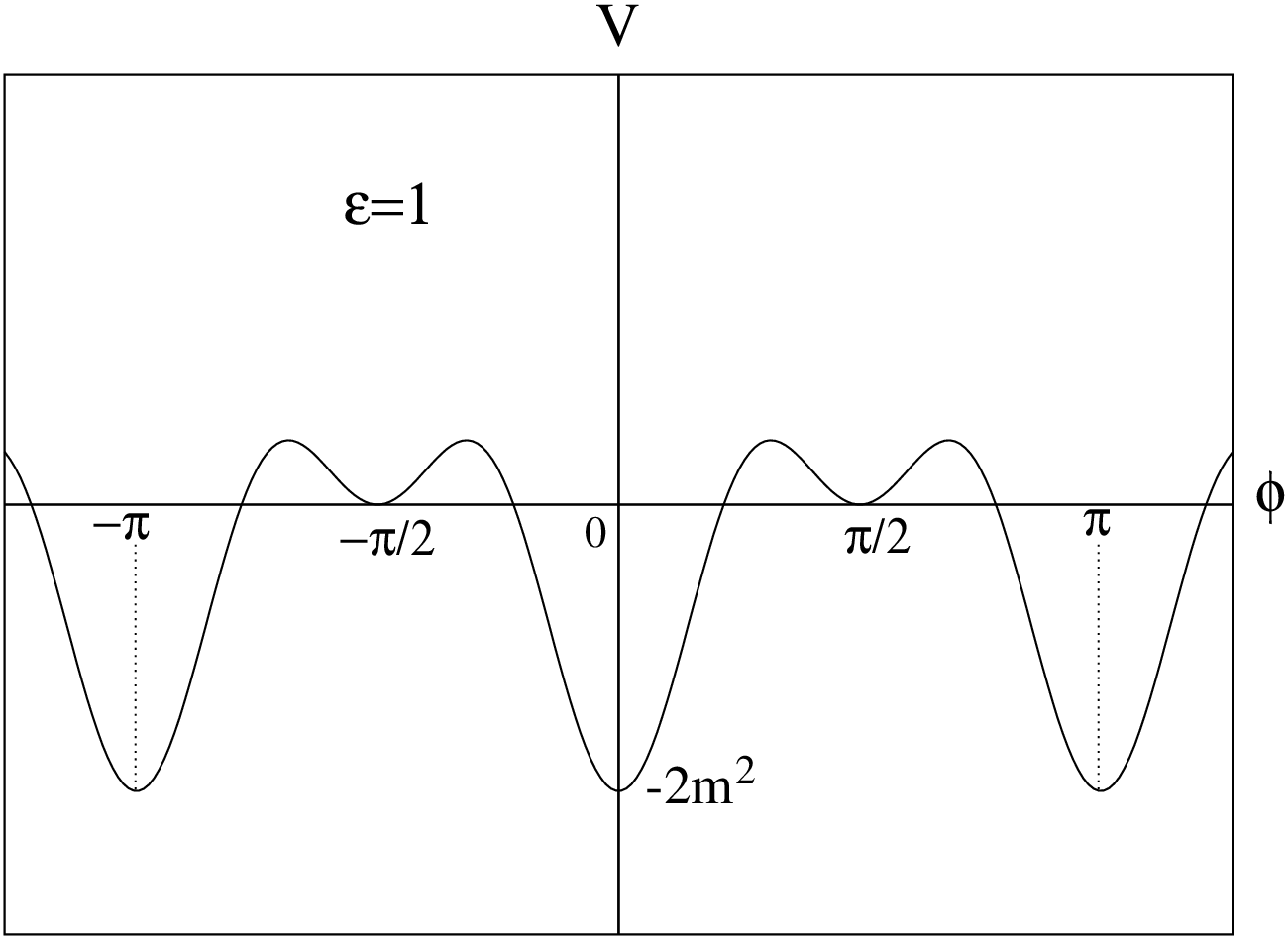}
}
\caption{The potential $V(\phi)$ plotted against $\phi$.
}\label{potfig}
\end{figure}

The equation (\ref{susy}) can be integrated easily and the 
result is (Figure 2): \footnote {Here we made a coordinate change 
$y \rightarrow -y$ for $\e=-1$ in order to be able to associate larger 
energies with 
increasing $y$. An integration constant $y_0$, which determines the 
location of the singularity, has been set to zero by 
using the translational invariance along the $y$ direction.}

\be
\phi= \begin{cases}
{\frac{1}{2} {\rm ln} {\left(1+e^{-4ma^2y} \over
{1-e^{-4ma^2y}} \right)} \hs{5} \e=-1, \ \ \ \ 0 \leq y < \infty \cr\cr
{\rm arctan} e^{4ma^2y} \hs{10} \e=1, \ \ \ -\infty < y < \infty}
\end{cases}
\label{asymp}
\ee

From this, $A$ and $W$ can be solved using (\ref{susy}) and (\ref{ap}): 

\bea
A &=& -{\frac {\e}{4a^2}} {\rm ln}(e^{-8\e ma^2y}+\e) , 
\label{inf} \w2  
W &=& {\frac{2m} {1+\e e^{8\e ma^2y}}}.
\eea

Now $v=e^A$ is the RG scale of the CFT. If we denote the operator 
that is associated with the field $\phi$ on the boundary by 
${\cal{O}}_{\phi}$, then its 
$\b$-function is defined as \cite{beta1, beta2, beta3}:

\be
\b= v\frac{d\phi}{dv}=-a^2\frac{1}{W}\frac{\del W}{\del \phi}
=\begin{cases}{-2a^2\rm{tanh}\phi \hs{4} \e=-1, \ \ \ 0 \leq \phi < \infty    
\cr \cr
\ \ 2a^2\rm{tan}\phi \hs{7} \e=1,  \ \ \ 0 \leq \phi \leq \pi/2}
\end{cases}
\label{beta}
\ee

\begin{figure}
\centerline{\epsfxsize=3.4truein
\epsffile{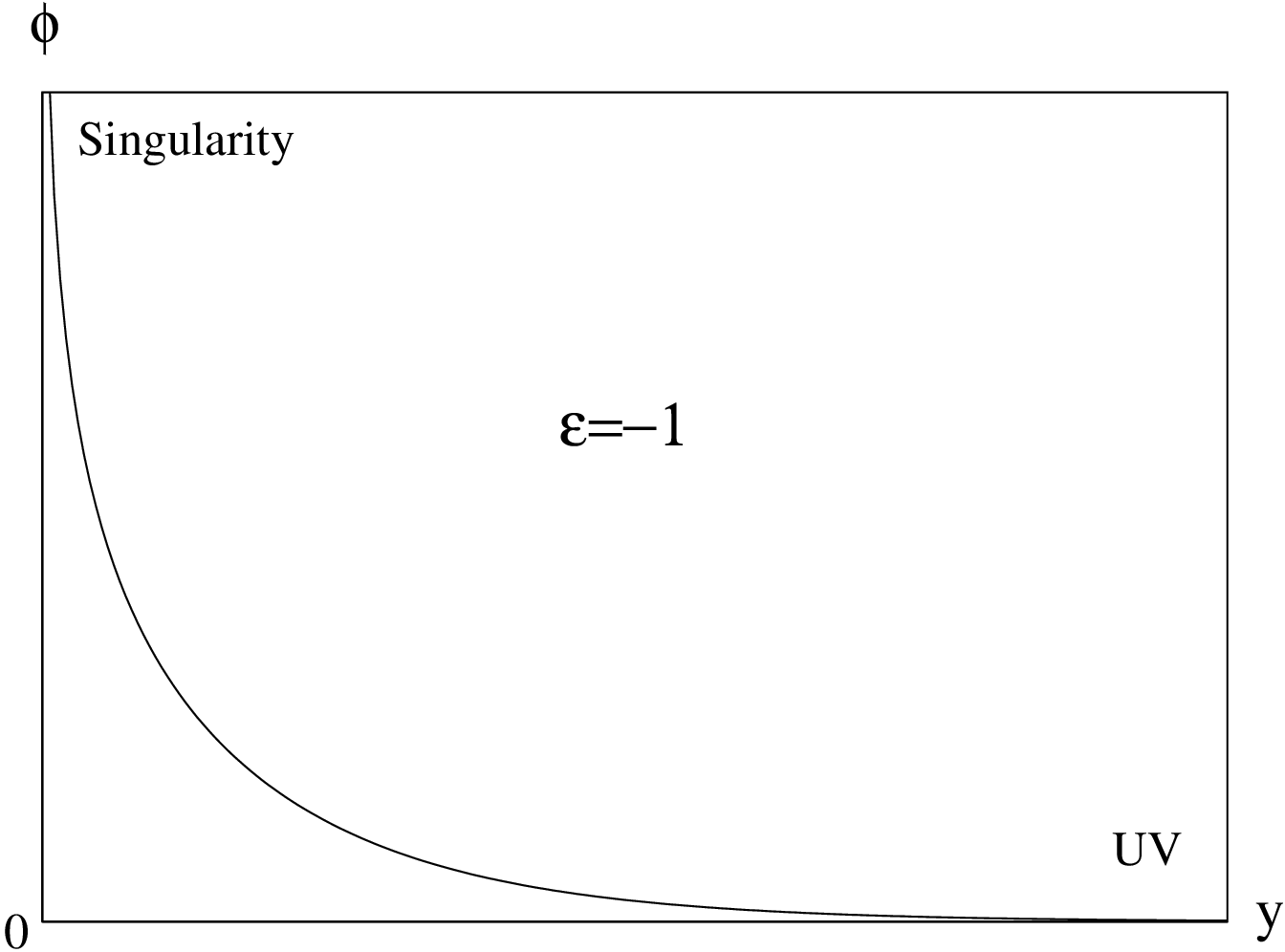}
\hspace{0.4in}
\epsfxsize=3.4truein
\epsffile{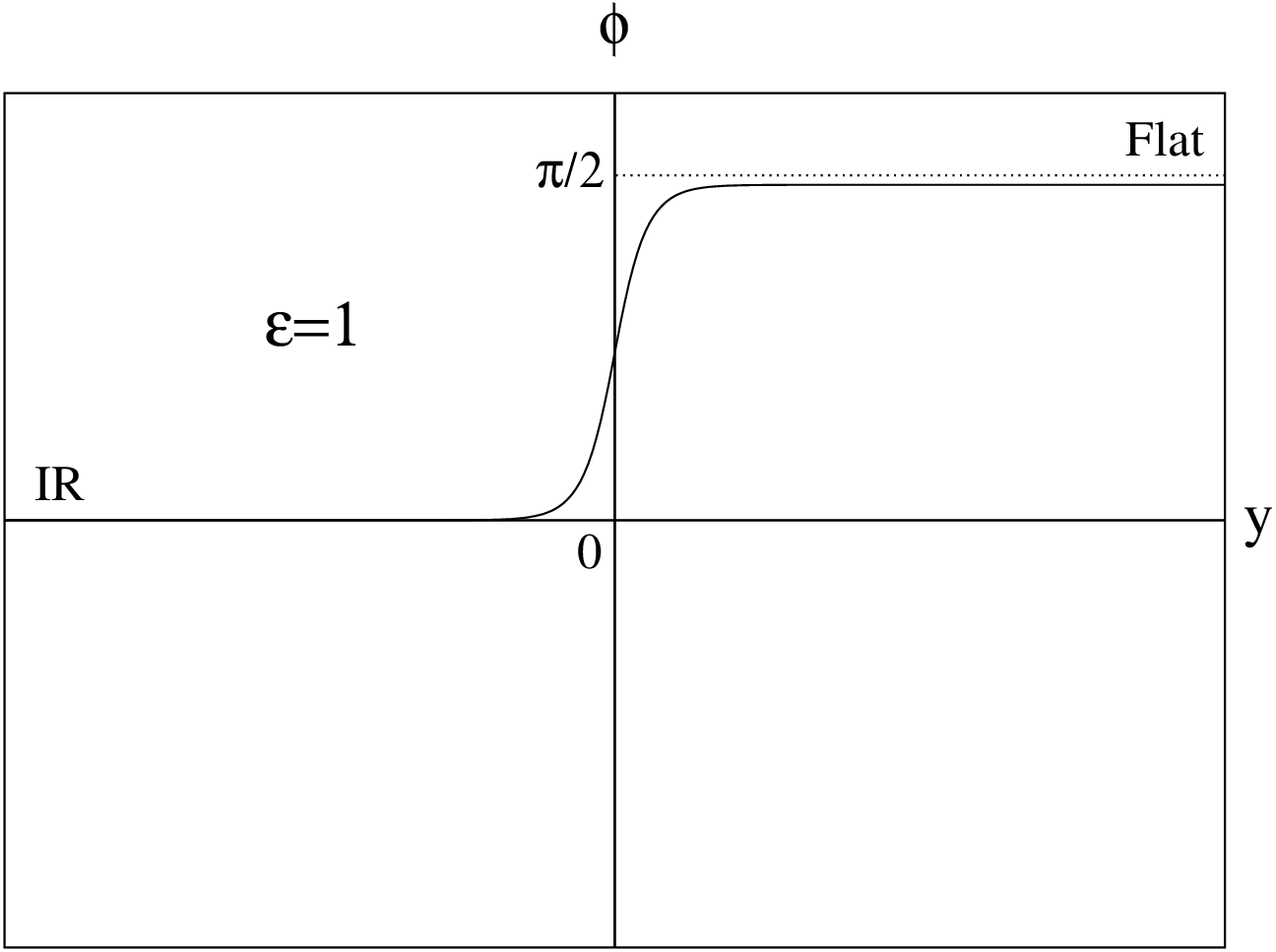}
}
\caption{The superymmetric solution $\phi$ versus $y$ graphs. For $\e=-1$ 
there is a 
naked 
singularity
at $y=0$ ($\phi \rightarrow \infty$) and an UV fixed point at 
$y=\infty$ ($\phi \rightarrow 0$).
For $\e$=1 there is an
IR fixed point at $y=-\infty$ ($\phi \rightarrow 0$). 
The spacetime becomes flat as
$y\rightarrow \infty$ ($\phi \rightarrow \pi/2$).
}
\label{figsusy}
\end{figure}

The sign of the derivative of the $\b$-function  
determines the nature of the fixed point at $\phi=0$. From (\ref{beta}) we 
deduce that
$\phi=0$ is an UV fixed point for $\e=-1$ and an IR fixed point for 
$\e=1$. 

Linearizing $V$ around $\phi = 0$, one finds that mass of the field $\phi$ 
at the 
fixed point is:
 
\be
M_{\phi}^{\, 2} = 16m^2a^2(a^2 +\e)
\label{mass}
\ee
 
From equations (\ref{linear}) we see that solution close to $AdS$ boundary
($\phi$=0) behaves as: {\footnote {In writing this equation we assume that 
$2 a^2 
\notin {\bf Z}$. Otherwise
there are some subtleties. See \cite{ads3} for a more detailed 
treatment.}}
 
\be
\phi =\phi_{0}^+ e^{-(2-\D) {y \over R}}
+\phi_{0}^-e^{-\D{y \over R}}
\label{bc}
\ee
 
$\D$ is the conformal dimension of the
boundary operator ${\cal{O}}_\phi$ and $R= {1 
\over 
2m}$ is
the $AdS$ radius. For a relevant operator $\D< 2$.  
From $\D(\D-2)= M_{\phi}^{\, 2}R^2$ and (\ref{mass}) we get 
$\D= 1+|1+2\e a^2|$. 
Now, let us analyze $\e=\pm$1 cases separately:

\subsection{Compact Sigma Manifold $S^2$ ($\e$=1)}

The potential \eq{v1} has a minimum at $\f=0$ which is a
supersymmetric $AdS_3$ vacuum and minima at ${\rm cos}\f=0$ corresponding  
to a
supersymmetric $2+1$ dimensional Minkowski vacua. The maxima at
${\rm cos}^2\f = a^2/(2a^2+1)$ are non-supersymmetric de 
Sitter
vacua (Figure 1). At these extrema scalars have 
tachyonic mass of $-32m^2a^4(a^2+1)/(2a^2 +1)$, so the de 
Sitter vacua are unstable.
The superpotential (\ref{superpot}) has two extrema which are located at 
${\rm cos}\f=1$ 
($AdS$ 
vacua) and
${\rm cos}\f=0$ (Minkowski vacua). This type of superpotential was 
also studied
in \cite{behrndt} for $N=2, D=5$ gauged supergravity. However, its results 
are general and apply to our case as well. 
As we mentioned following (\ref{beta}), $\phi=0$ ($y=-\infty$) 
is an IR fixed point. The dimension of the operator ${\cal{O}}_{\phi}$ is 
$\D=(2+2a^2) \geq 2$, and around $y=-\infty$, from (\ref{asymp}) we get 
$\phi\approx e^{-(2-\D)\frac{y}{R}}$ 
which 
means 
that we have a deformation of the CFT by an irrelevant operator 
\cite{rev3}.

\begin{figure}
\centerline{\epsfxsize=3.4truein
\epsffile{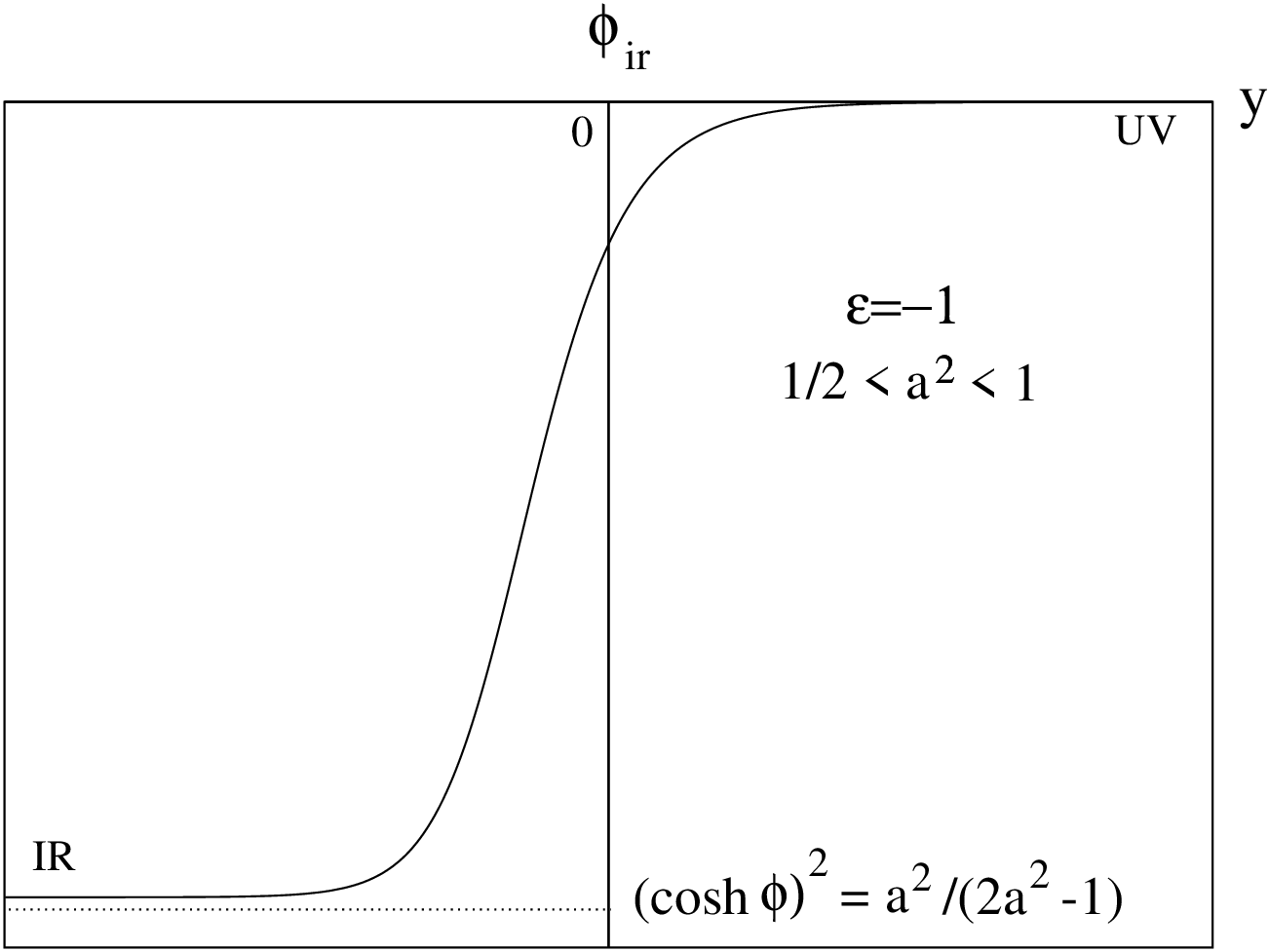}
\hspace{0.4in}
\epsfxsize=3.4truein
\epsffile{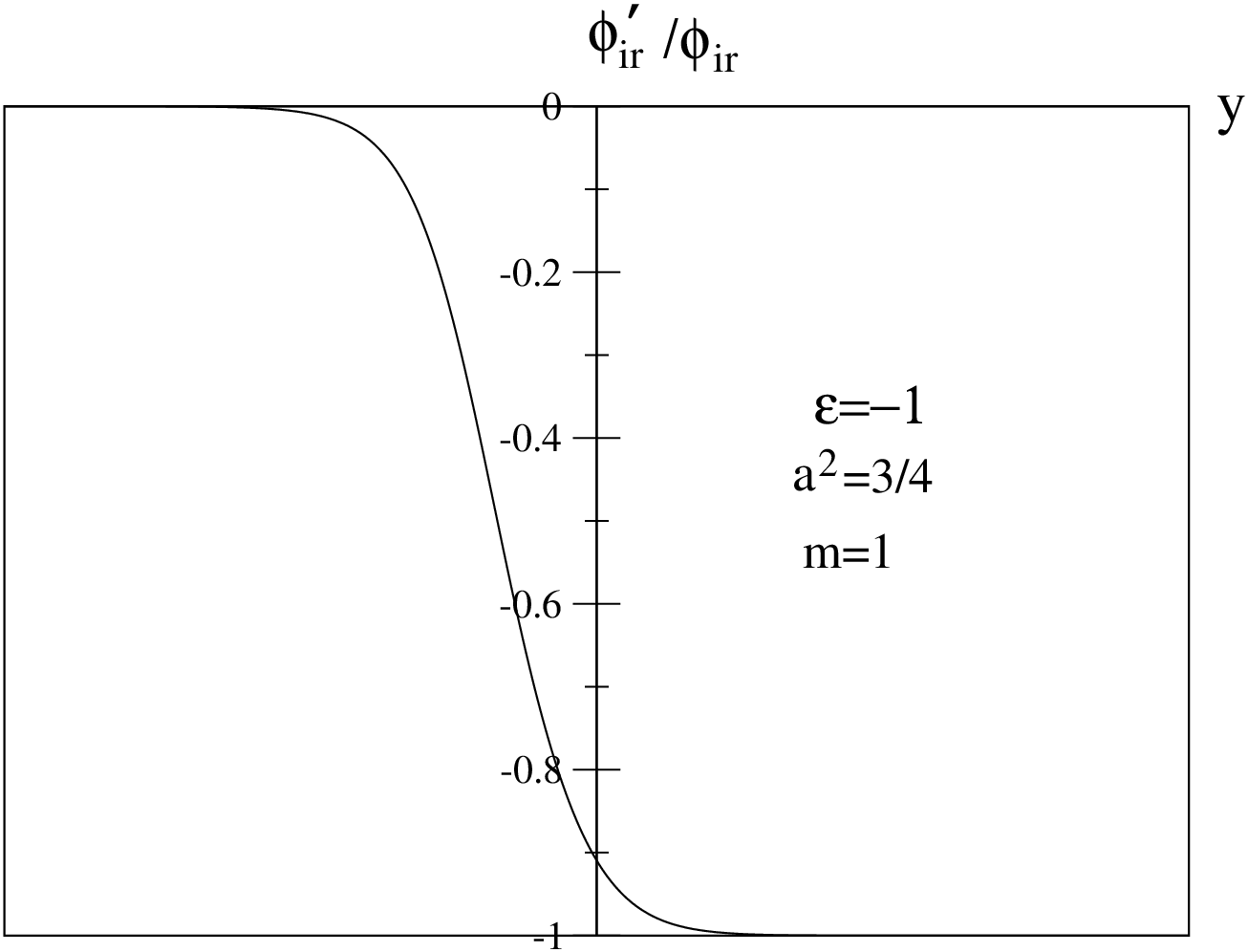}
}
\caption{The non-supersymmetric solution $\phi_{\rm ir}$.  
This solution interpolates between two $AdS$ vacua. 
$y=\infty$ is the UV and $y=-\infty$ is the IR region.
From the second graph we see
that
$\phi_{\rm ir} \approx e^{-(2-\D)\frac{y}{R}}= e^{-2m(2-2a^2)}$ as 
$y\rightarrow \infty$.}
\label{ir}
\end{figure}

\subsection{Non-compact Sigma Manifold $H^2$ ($\e$=-1)}

The theory admits various critical points (Figure 1):
 
(i)
For $a^2 \le 1/2$, the maximum at $\phi=0$ is a supersymmetric
$AdS_3$ vacuum.
 
(ii) For $1/2 < a^2 <1$, $\phi=0$ is a supersymmetric
$AdS_3$ vacua and
there are two 
minima at ${\rm cosh}^2\phi = {a^2 \over (2a^2-1)}$.
Although these are non-supersymmetric $AdS_3$ vacua, they are
stable  in the sense that they
satisfy the Breitenlohner-Freedman bound \cite{breit} ,i.e.,
${\del ^2 V \over \del \phi ^2} = 16m^2a^2{(1-a^2) \over (2a^2-1)}>-1$.
 
(iii) For $a^2 \ge 1$ the minimum at $\phi=0$ is
a supersymmetric $AdS_3$ vacuum.

Now, $y=\infty$ corresponds to an UV fixed point of the dual CFT. There is 
a 
naked singularity at $y=0$ (Figure 2). \footnote{In Figures (2)-(4) we 
assume,
without loss of generality,
$a^2= {3\over 4}$ and $m=1$.} This can be seen by looking at 
the metric (\ref{m}) close to $y=0$ using (\ref{inf}):

\be
ds^2=y^{{1 \over 2a^2}}(-dt^2+ dx^2) + dy^2 \, ,
\label{singular}
\ee

According to the analysis of \cite{gubser} this singularity is unphysical 
when $a^2 > 1/2$ despite the supersymmetry.\footnote{Exactly the same 
interval was excluded in 
\cite{ads3} since 
some fields become divergent on the 
boundary for $a^2 > 1/2$ and linear approximation applied in \cite{ads3} 
fails.} 
The reason is that, for $a^2 > 1/2$ the 
potential becomes unbounded near the singularity at $\phi=\infty$
as can be seen from (Figure 1). For $a^2\leq1/4$ the singularity is null 
and 
as explained in more detail in \cite{gubser} it is harder to understand 
the physics for $1/4<a^2\leq 1/2$ where the singularity is timelike.

In this 
supersymmetric solution the theory flows from an UV fixed point at $\phi 
=0$
($y=\infty$) to a non-conformal IR vacua at $\phi=\infty$. The rate which 
$\phi$ approaches to $y=\infty$ is $e^{-4ma^2y}$ from 
(\ref{asymp}).
Since in the same limit the determinant of the metric (\ref{m}) is $e^{2A} 
\approx
e^{4my}$ from (\ref{inf}), we see that $\phi$ is 
square-integrable when 
$a^2>1/2$ and therefore corresponds to giving a 
nonzero vacuum expectation value to the operator
$\mathcal{O}_\phi$.
For $a^2\leq1/2$ the solution can be interpreted as deformation 
by a relevant ($a^2=0$ case is marjinal) operator.
The form of the superpotential (\ref{superpot}) resembles the one 
studied in 
\cite{girardello} where their exact solution is associated with the RG 
flow from 
$N=4$ super Yang-Mills theory to pure $N=1$ in the IR by a deformation.

As we have discussed above, the supersymmetric solution does not 
physically make sense for $a^2>1/2$. However, 
for $1/2 < a^2 <1$ one can look for a 
kink solution that interpolates between the
maximum and a minimum \footnote{ Since the potential is  
symmetric around the y-axis it doesn't matter which minimum we choose. We 
will 
pick the one on the negative $\phi$-axis
for the forthcoming discussions.} of the potential (Figure 1). 
But, to do this one needs to solve 
the nonlinear field equations (\ref{linear}) 
with appropriate boundary conditions which is hard to 
do analytically. 
But, they can be solved numerically and the solution is given in (Figure 
3). \footnote {We follow the notation of \cite{seven} and use 
the
subscript ${ir}$ to
indicate the numerical solution. } From 
the second graph in (Figure 3) we see that, $\phi_{\rm ir}$
approaches
to $y=\infty$
as $e^{-2m(2-2a^2)y}$ and it is not square-integrable. Thus, this 
solution describes a relevant deformation the UV 
Lagrangian at $\phi=0$ by adding the term $\int\phi\mathcal{O}_\phi$. 
The 
theory flows to a non-supersymmetric IR fixed point at 
$\phi= {1 \over 2}\, {\rm ln}\left({1-2a\sqrt{1-a^2} \over 2a^2-1}\right)$
which is conformal 
again. At this extremum the reflection symmetry of the potential is broken  
and super-Higgs effect can be seen from (\ref{fermionic}).

\begin{figure}
\centerline{\epsfxsize=3.4truein
\epsffile{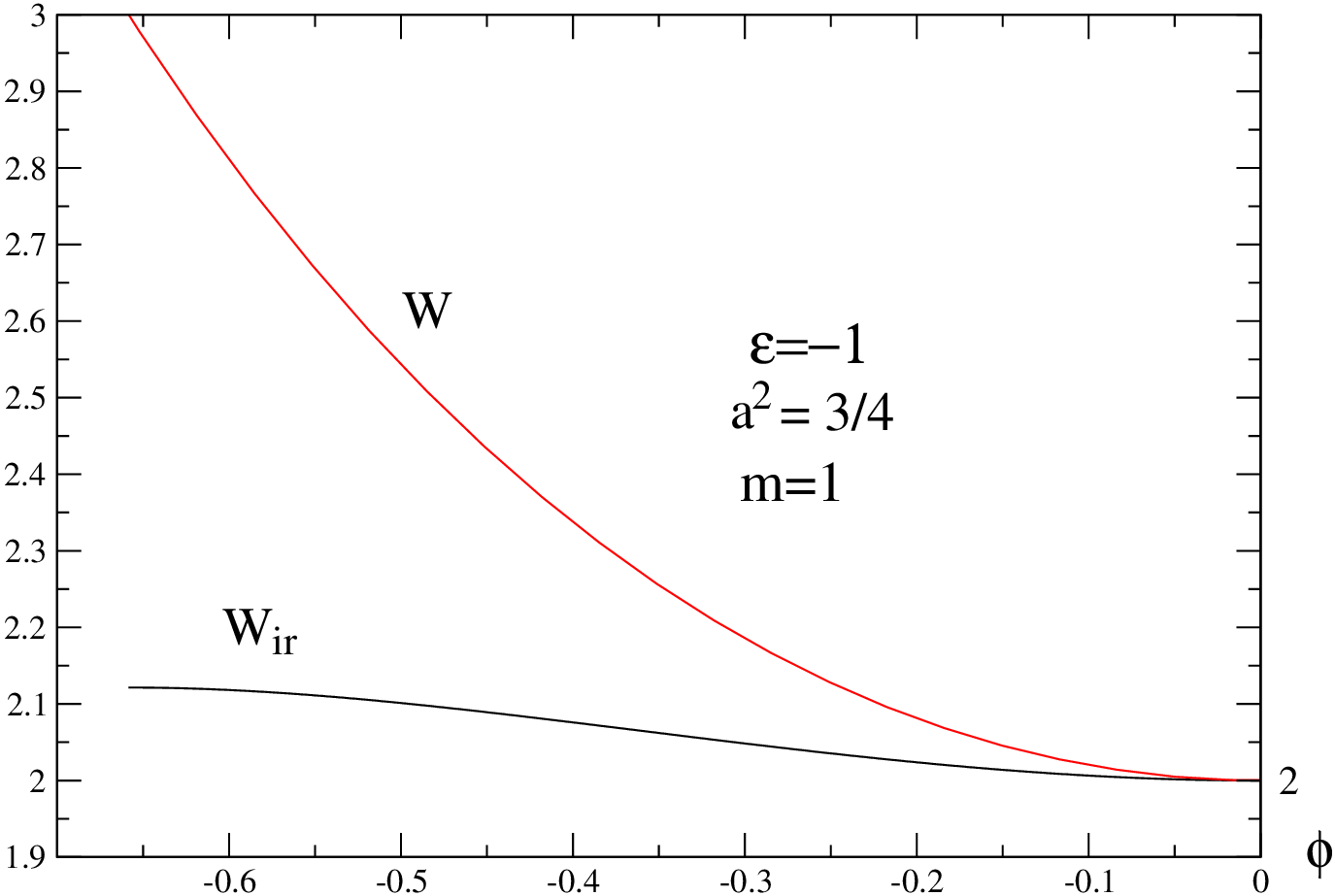}
\hspace{0.4in}
\epsfxsize=3.4truein
\epsffile{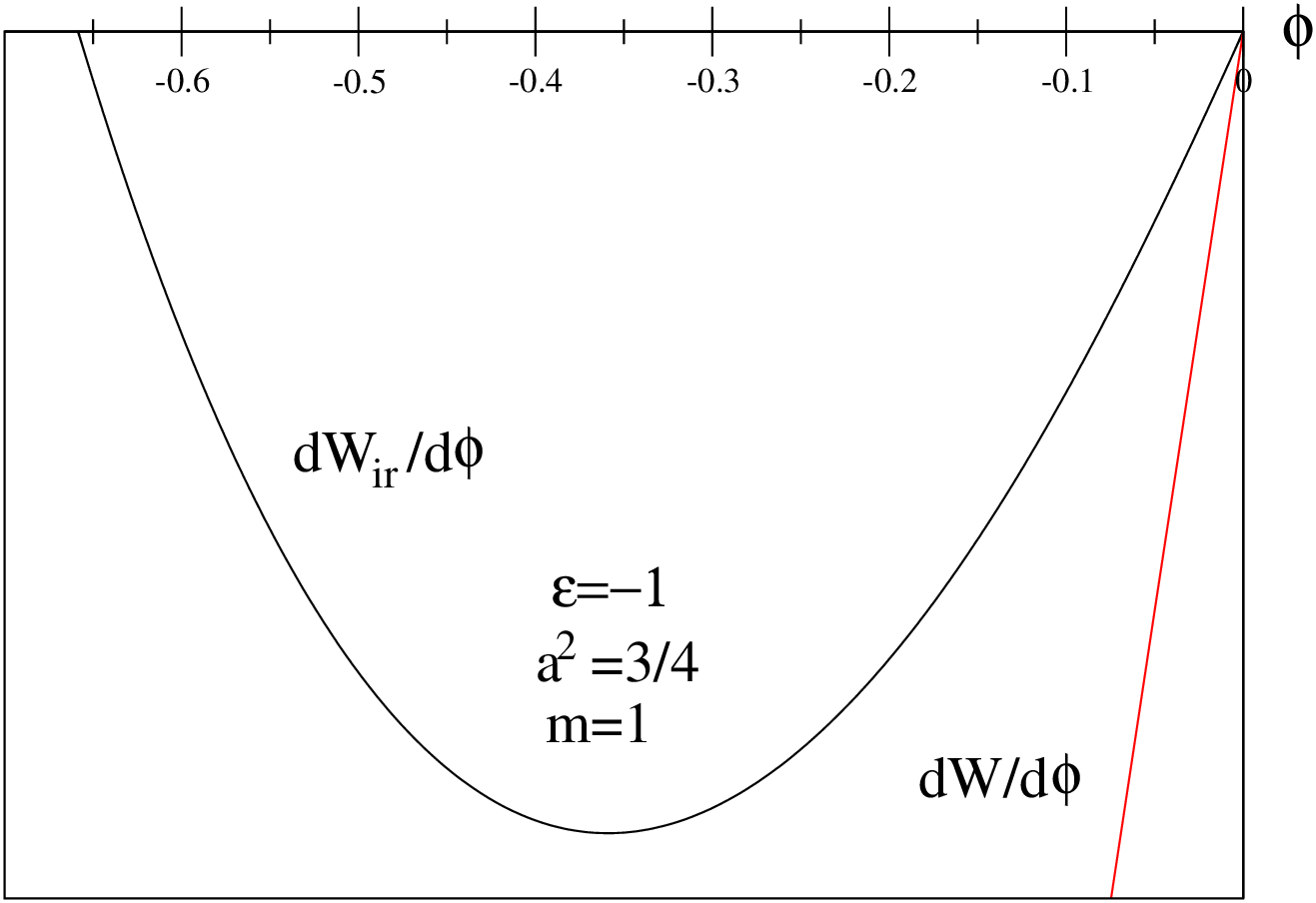}
}
\caption{The comparison of the generating function $W_{\rm ir}$ and 
the true superpotential $W$. $\del W_{\rm ir} \over \del \phi$ graph shows 
that
$W_{\rm ir}$ has a second extremum located at 
$\phi= {1 \over 2}\, {\rm ln}\left({1-2a\sqrt{1-a^2} \over 
2a^2-1}\right)$.} 
\label{figw}
\end{figure}

Actually any solution of (\ref{susy}) and (\ref{ap}) with $W$ 
satisfying the constraint (\ref{potential}) is a solution to second order 
equations \cite{skenderis1, free1}. 
$W$'s with this property depend on one integration constant and
only one of them is the true 
superpotential, the rest being the  Hamilton-Jacobi generating functions
\cite{beta2}. 
Once the 
solution for $\phi_{\rm ir}$ is known (Figure 3) the generating function 
$W_{\rm ir}$ 
can be 
constructed retrospectively \cite{seven} using equation (\ref{susy}) 
(Figure 4). 
In order to compare $W_{\rm ir}$ and the true superpotential $W$ we may 
make a Taylor 
series expansion \footnote{This expansion is studied in detail in 
\cite{martelli}. Our expansion agrees with theirs  after appropriate 
scalings.}  
around 
$\phi=0$ in (\ref{potential}). This is achieved by 
calculating 
the left-hand side of 
(\ref{potential}) from  (\ref{pot}) and (\ref{v1}) and 
then comparing with
the right-hand side term-by-term. Imposing $\del W/\del\phi =0$ at 
$\phi=0$ sets the coefficient of the term $\phi$ in $W$  to zero and we 
obtain a 
second order 
equation for the coefficient of $\phi^2$.
The result is 
$W_{\rm ir}\approx 2m(1+{(1-a^2)\over a^2}\phi^2 + \cdots)$, whereas
$W \approx 2m(1+ \phi^2 + \cdots)$. Even though $W$ 
has 
only 
one extremum which is at $\phi=0$, $W_{\rm ir}$ has two (Figure 4).

The holographic $\mathcal{C}$-function can be defined as
\cite{girar,free,beta3}:
 
\be
\mathcal{C} = { c_0\over A'}
\ee
 
where, $c_0$ is a constant. Since equations (\ref{linear})
imply that
$A''<0$, $\mathcal{C}$ decreases monotonically along the flow from
UV to IR. This function
should interpolate between the central charges $c_{\rm uv}$ and $c_{\rm
ir}$. From (\ref{potential}) and (\ref{ap}) we see that at fixed points
$A' \, ^2 = -2 V$ and thus,
 
\be
{c_{\rm ir} \over c_{\rm uv}}= {\sqrt{2a^2-1} \over a^2},
\hs{5} 1/2<a^2<1
\label{ratio}
\ee
 
Note that $c_{\rm ir}$ is less than $c_{\rm uv}$ as it should be by
Zamolodchikov's $\mathcal{C}$-theorem \cite{zamo}. The dimension of the 
operator at this extremum is $\D_{\rm ir}= 1+\sqrt{9-8a^2}$ 
from the potential (\ref{pot}). The ratio of the central 
charges is rational when $2a^2= r^2 +1$, where $r<1$ is a rational number.
If, in addition there exists a rational number $s$ such that $5=4r^2 
+s^2$, then operator dimensions in IR are
rational too, as happened in \cite{sam2}. \footnote{For matter 
fermions, $\l$, the
operator dimension is $\D_{\rm uv} = 1+{1 \over 2}|1-4a^2|$ and 
$\D_{\rm ir}= 1+{a^2 \over 2}{|1-4a^2| \over (2a^2-1)}$. 
\cite{ads3}} For example, $a^2=5/8$ satisfies both conditions.

\section{Spectrum of Massless Scalars}

In this section we solve the Laplace equation for a massless scalar 
field $\Phi$ in our supersymmetric background (\ref{m}). 
This problem has been investigated in higher dimensions first in 
\cite{free3, 
lap2, calogero, lap4}.
We begin by making a 
coordinate change $dy=e^{A}dz$ and obtain a conformally flat metric:

\be
ds^2=e^{2A(z)}(-dt^2+dx^2+dz^2)
\label{flat}
\ee

Then, the $\square_3\Phi=0$ equation with the above metric becomes: 
\footnote{If we make a small perturbation of the metric (\ref{flat}) as 
$ds^2=e^{2A}(\eta_{ij}+h_{ij})dx^idx^j +e^{2A}dz^2$
it can be shown that the transverse, traceless part of $h_{ij}$
obey the same wave 
equation (\ref{box}) as happened in higher dimensions
\cite{sfetsos}. 
However, in 3 dimensions graviton has no degree of freedom 
and imposing $h^i_i=\partial^ih_{ij}=0$ automatically implies 
$(-\partial_t^2 +\partial_x^2)h_{ij}=0$. This suggests that scalar 
fluctuations should also 
be included to analyze metric fluctuations.}

\be
(\partial_z^2 +\partial_z A\partial_z -\partial_t^2 
+\partial_x^2)\Phi(t,x,z)=0
\label{box}
\ee

After making a separation 
of variable,
$\Phi(t,x,z)=e^{i\vec{p}.\vec{x}}e^{-A/2}\Psi(z)$ we get,

\be
\partial_z^2\Psi -V_{QM}\Psi=p^2\Psi
\label{box2}
\ee

where the potential is given by,

\be
V_{QM}=\frac{1}{2}\partial_z^2 A + \frac{1}{4} (\partial_z A)^2
\ee

Note that the potential can be written as 
$V_{QM}={\cal{U}}(z)^2 + {\cal{U}}'(z)$ where 
the 
prepotential is
$2{\cal{U}}(z)=\del_zA$ and hence this is a supersymmetric quantum 
mechanics problem.
This implies the positivity of the normalizable 
spectrum \cite{wit1,free1,free2}. In other words, the physical spectrum is 
bounded 
from below, $-p^2\geq 0$.
It is usually enough to know the 
shape of  
$V_{QM}$ to get some 
qualitative information about the dual CFT spectrum \cite{free3}. 
To accomplish this, it is sufficient to obtain coordinate $z$ in terms of    
$y$ near the end points from (\ref{inf}). 
For 
$\e=-1$
the $AdS$ boundary, i.e. $y=\infty$, is mapped into $z=z_0$ and the 
potential blows up quadratically, 

\be
V_{QM}(AdS) \approx \frac{3}{4(z-z_0)^2}
\label{limit}
\ee

whereas the potential close to the singularity at $y=0$ is 

\be
V_{QM}(\rm{singularity}) \approx \frac{(3-8a^2)}{\,\,\ 4(1-4a^2)^2 \ z^2}. 
\ee 

The singularity located at $y=0$ is mapped into $z=0$ and and $z=-\infty$
for $a^2>1/4$ and $a^2\leq 1/4$ respectively. Therefore, for $a^2\leq1/4$ 
the 
potential approaches to a finite 
value and we have a
continuous spectrum. Only $a^2=1/4$ case is unclear because of the 
denominator, which implies a mass gap 
as 
we will see below. 
When $a^2>1/4$ in order to have a discrete spectrum the 
coefficient of 
the potential should satisfy $(3-8a^2)/4(1-4a^2)^2\geq -1/4$ from 
elementary quantum mechanics. This is indeed 
true for any $a^2$. The equality occurs at $a^2=1/2$. 
The potential can be solved exactly for $\e=-1$ when $a^2=1/2$ and 
$a^2=1/4$:

\begin{figure}
\centerline{\epsfxsize=2.3truein
\epsffile{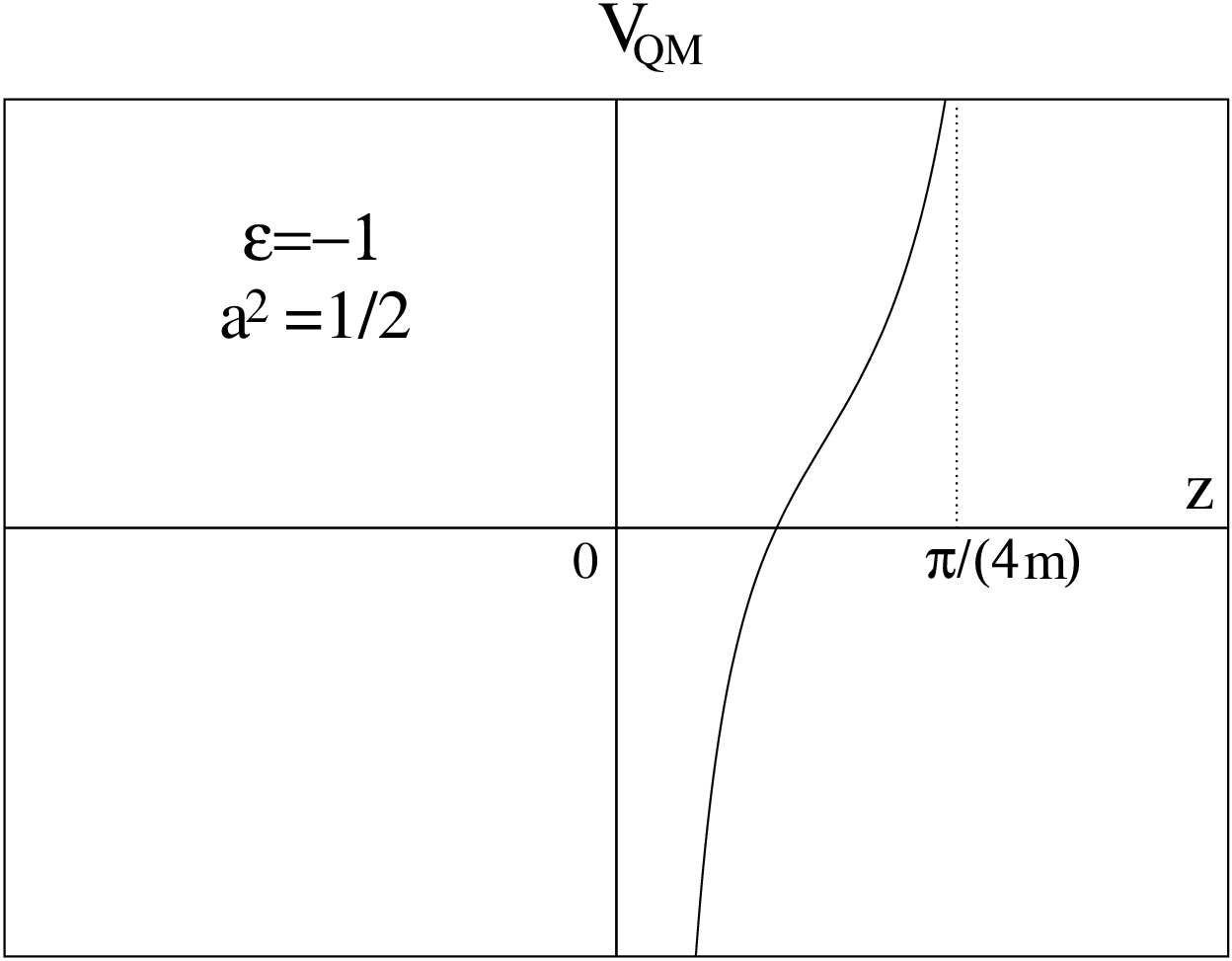}
\hspace{0.25in}
\epsfxsize=2.3truein
\epsffile{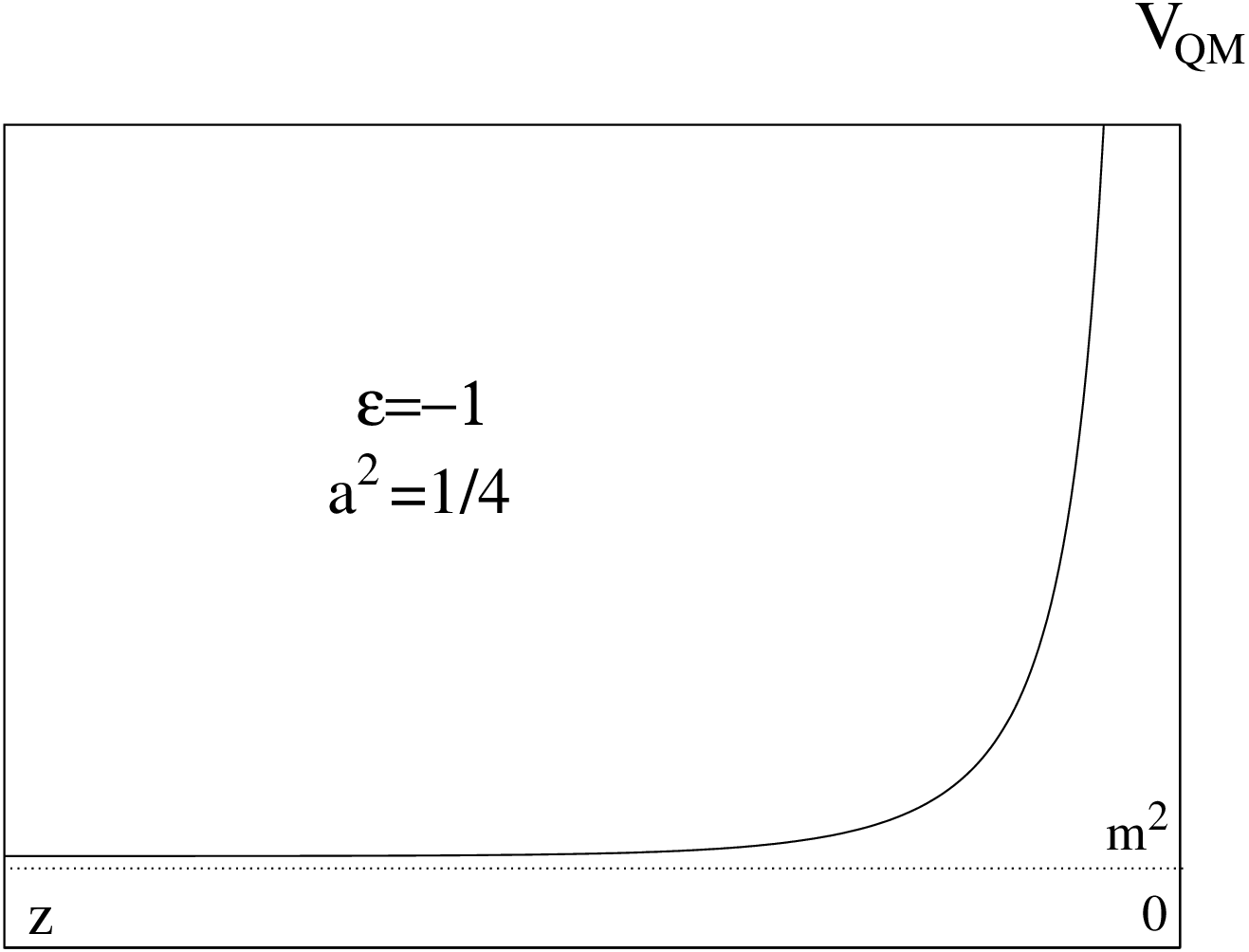}
\hspace{0.25in}
\epsfxsize=2.3truein
\epsffile{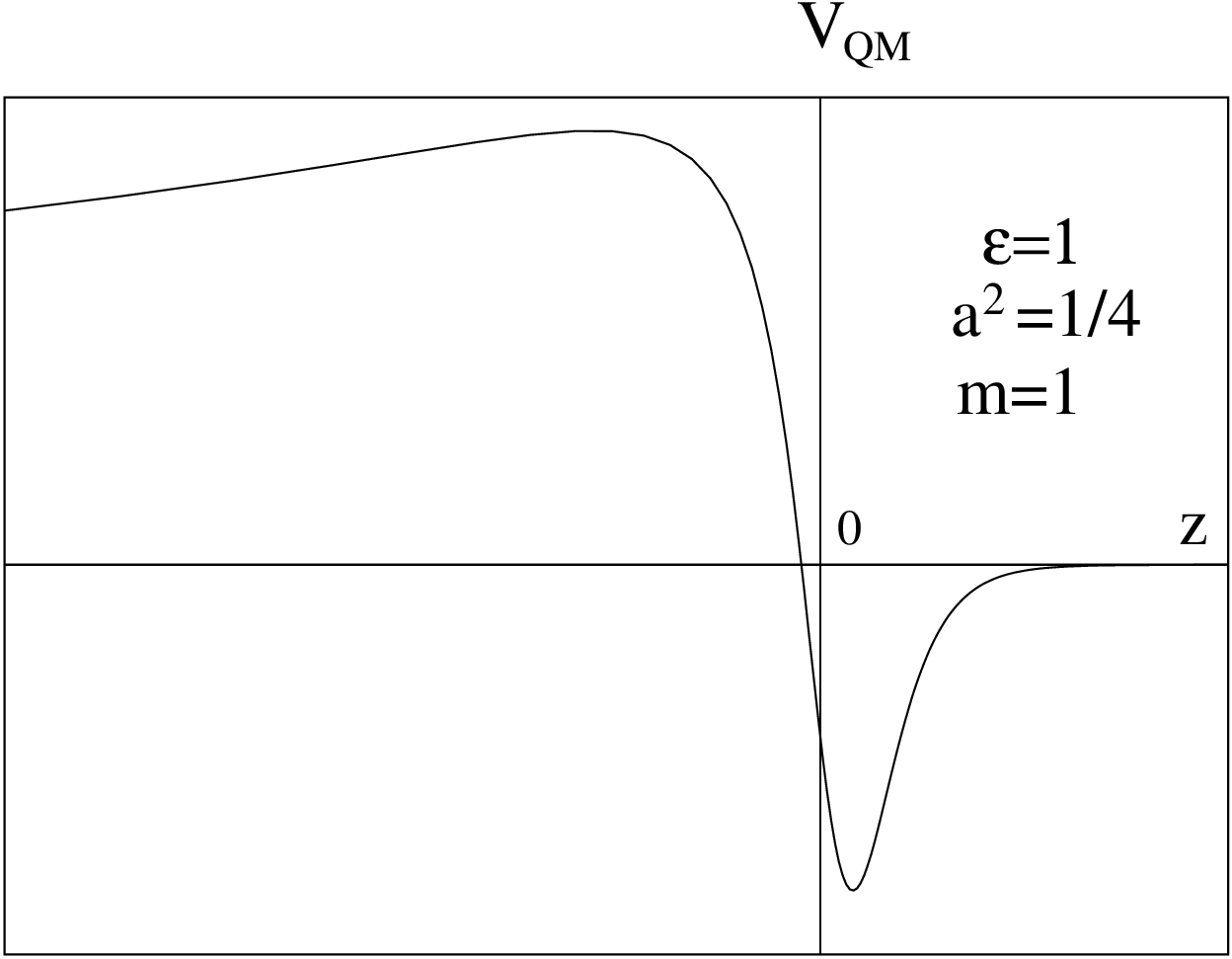}
}
\caption{Quantum mechanical potentials. For $\e=-1$ and $a^2\leq 1/4$, 
$V_{QM} \rightarrow 
0$ as $z \rightarrow -\infty$, except when $a^2=1/4$.}
\end{figure}

\be
V_{QM}=\begin{cases}{\frac{4m^2(1-2{\rm cos4mz})}{{\rm sin^24mz}} \hs{5}
a^2=1/2, \ \ \ 0\leq z<\pi /4m \cr\cr
\frac{2m^2e^{2mz}+ m^2}{(1-e^{2mz})^2} \hs{9} a^2=1/4,\ \ \ 0<z<-\infty}
\end{cases}
\label{vqm}
\ee

When $a^2=1/4$ we see that $V_{QM} \rightarrow m^2$ as $z \rightarrow 
-\infty$ which implies a mass gap (Figure 5). These potentials are among 
the ones studied in Calogero models as was noticed in 
\cite{calogero} for $AdS_5$. 
The first one is a member of the 
P\"{o}schl-Teller potentials (type I) and the second one belongs to 
Eckart type potentials. (See \cite{qm} for a review.)

For $\e=1$ the $AdS$ boundary at $y=-\infty$ is mapped into $z=-\infty$ 
and 
the potential is given by (\ref{limit}). The flat region at $y=\infty$ is 
mapped into $z=\infty$ and $V_{QM}\rightarrow 0$ in this limit too. 
Therefore we have a continuous spectrum without any mass gap. 
Unfortunately we 
couldn't solve the potential in a closed form for any value of $a^2$.
However, it can be drawn numerically; for instance, when $a^2=1/4$
it looks like in (Figure 5).

\subsection{2-point functions ($\e=-1$)}

We may  check the above statements explicitly when $\e=-1$  for $a^2=1/2$ 
and $a^2=1/4$
by calculating the 2-point functions. 
For $a^2=1/2$ let us define a new 
variable $u=\rm{sin}^2 (z/R)$ where 
$R=\frac{1}{2m}$. Now the singularity is at $u=0$ and the $AdS$ boundary 
is 
at $u=1$. The equation (\ref{box2}) becomes:

\be
u(1-u)\Phi''+(1-u)\Phi' - \frac{p^2R^2}{4}\Phi =0
\label{hyper1}
\ee

This is a hypergeometric equation and its solution which is regular in the 
interior ($u=0$) is:

\be
\Phi=F\left(\sqrt{-p^2R^2\over{4}}, -\sqrt{-p^2R^2\over{4}}; 1 ; u\right)
\ee

The other linearly independent solution of (\ref{hyper1}) is 
logarithmically divergent at $u=0$.
Now, we can calculate the 
2-point 
function using the prescription of \cite{mal2,cor}  and get:{\footnote{We 
have used equations (15.1.20), (15.2.1), (15.3.10), (15.3.11) of 
\cite{math}
and some properties 
of the $\psi$-function in deriving (\ref{cor1}) and (\ref{cor2}).}

\be
\left< {\cal{O}} (p) {\cal{O}} (-p) \right> = -\frac{p^2R^2}{4}
\left[ \psi\left(1+\sqrt{-p^2R^2\over{4}} \right) +
\psi\left(1-\sqrt{-p^2R^2\over{4}} \right) \right]
\label{cor1}
\ee

where $\psi=\Gamma '/\Gamma$ is the psi-function.
The correlator has a discrete spectrum with poles located at 
$-p^2=4(n+1)^2/R^2 , \ \ n=0,1,2,...$. Note that there is no 
zero-mass pole.

For $a^2=1/4$ we define a new variable $u=e^{z/R}$ and the equation 
(\ref{box2}) becomes:

\be
u^2(1-u)\Phi'' + u(1-u)\Phi' - p^2R^2(1-u)\Phi=0
\label{hyper2}
\ee

This is again a hypergeometric equation and its regular solution at 
$u=0$ is:

\be
\Phi=u^{-\frac{1}{2}+q} 
F\left(-\frac{1}{2}+q +\sqrt{p^2R^2}, -\frac{1}{2}+q -\sqrt{p^2R^2} ; 
1+2q ; u\right) , 
\hs{5} 
q= \frac{1}{2}\sqrt{1+4p^2R^2}
\ee

The other solution of (\ref{hyper2}) has a leading 
$u^{-\frac{1}{2}-q}$ factor and therefore it is irregular. The 2-point 
correlator up 
to a normalization is:

\be
\left< {\cal{O}} (p) {\cal{O}} (-p) \right> = -p^2R^2
\left[ \psi\left(\frac{3}{2}+q+\sqrt{p^2R^2}\right) +
\psi\left(\frac{3}{2}+q-\sqrt{p^2R^2}\right) \right]
\label{cor2}
\ee

There is a branch cut along the real $p^2$-axis that extends over the 
interval 
$(-\frac{1}{4R^2}, -\infty)$. The spectrum is 
continuous with a mass gap of $M_{gap}^2=\frac{1}{4R^2}=m^2$ 
as we 
observed 
above. 
Note that $q$ 
is purely imaginary after this gap.

Both of the correlation functions (\ref{cor1}) and (\ref{cor2}) for large 
timelike momentums behave as $-p^2\rm{ln}p$ which is the pure $AdS$ form.


\section{ Conclusions }


In this section we would like to indicate some open problems
and possible applications of the model we studied. This simple system
can be a useful toy model for understanding some complicated problems that
arise in higher dimensions. For example, when the sigma model manifold is
compact, close to the Minkowski vacuum the cosmological constant is
positive and gets smaller as we approach to the extremum (Figure 1). As
mentioned in \cite{behrndt}, this might be convenient for 
studying quintessence 
scenario \cite{quint}. 
If $dS$/CFT correspondence \cite{ds} is realized, then the 
holographic duals of Euclidean $AdS$ space and $dS$ space may be related 
to 
each other \cite{boer1}. The de Sitter vacua that appears when 
$\e=1$, might be used to explore this connection in supergravity.(See also
\cite{argurio} for some comments.)

An important problem is to find the M-theory origin of this model. In fact 
this is a general problem for most of the known 3-dimensional  gauged 
supergravities because 
for vector fields there is a Chern-Simons term instead of the usual 
kinetic term in the Lagrangian \cite{ads2,nic1, nic2}.   
A possible solution might 
be reduction with background fluxes \cite{massive}.

After finding the compactification, the next step is to identify the CFT
dual of this model. To do this one has to figure out the brane
configuration that gives rise to that particular near-horizon geometry and 
then
find its world-volume description. This would allow us to compare  the 
results of this paper
such as he ratio (\ref{ratio}) and massless scalar spectrums, with the 
CFT.

It would be nice to understand the constant
$a^2$ both from the field theory and M-theory point of view. 
The naked singularity appeared in the
supersymmetric solution when $\e=-1$ is acceptable for $a^2\leq 1/2$
by the criteria given in \cite{gubser}. It is reasonable to expect
that the dimensional reduction will fix the value of $a^2$
from this interval. In order to explain why $a^2>1/2$ is excluded, the 
solution has to be lifted to M-theory. It may correspond to brane 
distributions with negative charge and tension \cite{free3, calogero, 
lap4}. When $\e=-1$, for $a^2=1/2$ we found
a discrete spectrum with $V_{QM}(z)=-1/(4z^2)$ and because of that 
$a^2=1/2$
may distinguish itself 
as happened in 5-dimensions \cite{rev1,gubser}. When $\e=1$, 
$a^2=1/2$ is again special since the black string solution 
(\ref{inf}) coincides with 
the one 
found in \cite{horowitz} which is a solution of the 2+1 dimensional string 
theory \cite{ads2}. Clearly, nothing much can be said without determining 
the higher dimensional origin.

The singularity that is present when $\e=-1$ is probably due to some 
strong coupling 
effect
in the dual CFT \cite{gir,lap2,gubser}. Wilson loops should be calculated
to see whether there is a confinement or screening \cite{wit1}  
phenomena in the field theory.

A natural continuation of the present work is to consider fluctuations of
fields other than massless scalars such as vector field, fermions, active 
and inert 
scalars, 
and apply the holographic 
renormalization techniques (see \cite{skenderis2} for a review and 
\cite{sam2, sam3} for its use in $D=3$) 
to 
obtain counterterms and correlators. We hope to do these in the near 
future.

\subsection*{Acknowledgements}
\hs{4} I am grateful to Tonguc Rador for his help in computer to obtain 
the 
numerical 
solution given in section 2.2. I also would like to thank Ergin 
Sezgin, Kostas Skenderis and especially to Ali Kaya for 
useful discussions.


\end{document}